\documentclass[journal,article,submit,moreauthors,pdftex]{Definitions/mdpi} 

\firstpage{1} 
\pubvolume{xx}
\issuenum{1}
\articlenumber{5}
\pubyear{2020}
\copyrightyear{2020}
\history{Received: date; Accepted: date; Published: date}

\usepackage{multicol}
\usepackage{multirow}
\usepackage{color}
\usepackage{xcolor}
\usepackage{booktabs}
\usepackage{graphicx}
\usepackage{nicefrac}
\usepackage{amsmath}
\usepackage{array}
\newcolumntype{P}[1]{>{\centering\arraybackslash}p{#1}}

\newtheorem*{remark}{Remark}

\usepackage{graphicx}
\usepackage{fancyhdr}
\usepackage{amssymb}
\usepackage{url}
\usepackage{color}
\usepackage{tikz}
\usetikzlibrary{positioning}
\usepackage{pgfplots}
\usepackage{marvosym}

\usepackage{float}
\usepackage{todonotes}

\usepackage{tikz}
\usepackage{tikzscale}
\usepackage{xspace}
\usepackage{pgfplots}
\usepgfplotslibrary{groupplots}
\pgfplotsset{compat=1.16}
\usepackage{xparse}
\NewDocumentCommand{\Log}{o}{%
  \IfNoValueTF{#1}{}{{}^{#1}\!}\log}%

\usetikzlibrary{external}
\Title{PWM2Vec: An Efficient Embedding Approach For Viral Host Specification From Coronavirus Spike Sequences}

\Author{Sarwan Ali $^{1}$,Babatunde Bello $^{1}$,Prakash Chourasia $^{1}$,Ria Thazhe Punathil $^{1}$, Yijing Zhou $^{1}$, Murray Patterson $^{2}$*}

\AuthorNames{Sarwan Ali,Babatunde Bello,Prakash Chourasia,Ria Thazhe Punathil, Yijing Zhou, Murray Patterson}
\address{%
$^{1}$ \quad Department of Computer Science, Georgia State University, Atlanta, GA, USA \\
\{sali85,bbello1,pchourasia1,rthazhepunathil1,yzhou43\}@student.gsu.edu
\\
$^{2}$ \quad Department of Computer Science, Georgia State University, Atlanta, GA, USA 
}
\corres{mpatterson30@gsu.edu}

\abstract{The origin of SARS-CoV-2 in humans, which led to the
  COVID-19 pandemic, is still unknown and is an important open
  question.  There are speculations that bats are a possible origin.
  Likewise, there are many closely related (corona-) viruses, such as
  SARS, which was found to be transmitted through civets.  The study
  of the different hosts which can be potential carriers and
  transmitters of deadly viruses to humans is crucial to
  understanding, mitigating and preventing current and future
  pandemics.  In coronaviruses, the surface (S) protein, or spike
  protein, is an important part of determining host specificity since
  it is the point of contact between the virus and the host cell
  membrane.
  In this paper, we classify the hosts of over five thousand
  coronaviruses from their spike protein sequences, segregating them
  into clusters of distinct hosts among avians, bats, camels, swines,
  humans and weasels, to name a few.  We propose a feature embedding
  based on the well-known position-weight matrix (PWM), which we call
  PWM2Vec, and use to generate feature vectors from the spike protein
  sequences of these coronaviruses.  While our embedding is inspired
  by the success of PWMs in biological applications such as
  determining protein function, or identifying transcription factor
  binding sites, we are the first (to the best of our knowledge) to
  use PWMs in the context of host classification from viral sequences
  to generate a fixed-length feature vector representation. The
  results on the real world data show that in using PWM2Vec, the
  machine learning classifiers are able to perform comparably well as
  the baseline models in terms of predictive
  performance and runtime --- in some cases the performance is better.
  We also measure the importance of different amino acids using
  information gain to show the amino acids which are important for
  predicting the host of a given coronavirus.  Finally, we perform
  some statistical analyses on these results to show that our
  embedding is more compact than the baselines.}

\begin{document}


\section{Introduction}
The coronavirus disease 2019 (COVID-19) pandemic is caused by the severe acute respiratory syndrome coronavirus 2 (SARS-CoV-2). The pandemic has put millions of people at risk in numerous countries worldwide and made an unprecedented public health crisis \cite{majumder2021recent}.
Although the origin of COVID-19 (SARS-CoV-2) in humans is still unknown, there are many theories that it could have been transferred to humans from bats~\cite{zhou-2020-pneumonia}.  Likewise, several
related coronaviruses (CoVs), such as SARS (SARS-CoV), was transmitted to
humans from civets (civets are closely
  related to cats~\cite{johnson-2006-radiation}), and MERS (MERS-CoV)
from dromedary camels~\cite{reusken-2014-mers}.
SARS-CoV-2, and other CoVs, belong to the family coronaviridae (of order nidovirales~\cite{order-nidovirales}), which is a large family of diverse enveloped, positive-sense single-stranded genomic RNA(+ssRNA) viruses that can
bring about respiratory diseases to humans and animals \cite{yuce2021covid}.
They are grouped into five genera namely alphacoronavirus,
betacoronavirus, gammacoronavirus, alphaletovirus and
deltacoronavirus.  They infect a range of hosts such as human, palm
civets, bats, dogs, monkeys among others~\cite{li2006animal}.
Alphacoronaviruses and betacoronaviruses largely infect mammals, and
gammacoronaviruses largely infect avian, the delta coronavirus both
avian and mammals~\cite{li2016structure} (see
Figure~\ref{fig_genus}).

SARS-CoV-2 is the seventh of this coronavirus family known to affect humans, and the other six are severe acute respiratory syndrome-CoV (SARS-CoV), HCoVs-NL63, HCoVs-OC43, HCoVs-HKU1, HCoVs-229E, and middle east respiratory syndrome-CoV (MERS-CoV) \cite{mungroo2021increasing}. SARS-CoV-2 is similar to SARS-CoV which is one of the previous virus from the family coronaviridae, that lead to the SARS epidemic in 2003 caused more than 8400 cases and approximately 800 deaths \cite{satija2007molecular}. As compared to the known SARS-CoV virus, the novel SARS-CoV-2 has a lower mortality rate but higher human-to-human transmission rate, also SARS-CoV-2 can have adverse impact on the human body. It is highly infectious and is the matter of big concern since it can not only damage the respiratory system, gastrointestinal system, heart, and central nervous system, but also may lead to multi-organ failure and eventually death \cite{li2020asymptomatic, guan2020clinical}.  

Monitoring zoonotic disease and host specificity are integral to understanding disease dynamics. The
60\% of known infectious diseases in humans and 75\% of  all emerging disease are zoonotic as per report by United Nations Environment Program (UNEP) and International Livestock Research Institute (ILRI)~\cite{UNEP_Report}.
The study of the COVID-19 pandemic is of great significance, not only because it can help healthcare institutions to cope with the ongoing epidemic but also because it allows researchers to learn more fundamentals about the family coronaviridae, which can provide new knowledge for the prevention of potential pandemics in the future. CoVs are widespread among birds and mammals and can be a cause for zoonoses. Zoonoses is defined as any disease or infection that is naturally transmissible from vertebrate animals to humans by the World Health Organization, and COVID-19 has been classified as a zoonotic disease~\cite{haider2020covid}. One important step to learn zoonoses and understand the pandemic better is finding how human infection began. CoVs can lead to various diseases in domestic animals, including dogs, pigs, chickens, and cats. Although the origin of COVID-19 in humans is still unknown, genetic analysis results show that it’s highly possible that SARS-CoV-2 originates from bats and utilizes the pangolin as an intermediate host~\cite{salian2021covid, han2020pangolins,  li2021epidemiological}.

\begin{figure}[!ht]
  \centering
  \includegraphics[scale = 0.5]{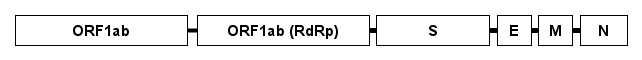}
  \caption{The genome of a coronavirus ranges from 26--32kb in length~\cite{order-nidovirales},
    each of them coding for two non-structural and four structural
    proteins.  The non-structural proteins are coded by open reading
    frame (ORF) 1a and 1ab, where ORF1ab contains the RNA-dependent
    RNA polymerase gene (RdRp).  The structural genes include spike
    (S), envelope (E), membrane (M) and nucleocapsid (N).  The S gene
    region encodes the spike protein, which is responsible for
    attaching the virus to receptors on the host cell membrane.}
  \label{fig_spike_seq_example}
\end{figure}

The CoVs have an envelope membrane that is associated with five structural proteins, namely the surface (S) protein, or spike protein, 
haemagglutinin-esterase protein (HE), membrane protein (M), envelope protein (E), and the nucleocapsid protein (N) \cite{umakanthan2020origin}, see Figure~\ref{fig_spike_seq_example}.
The spike protein mediates the binding and fusion between the virus and the host cell receptors, and also the infected host cells and adjacent uninfected cells \cite{rosales2020does}.
Hence the spike protein
of different CoVs largely
determines the range of host specificity of these CoVs.  Changes in spike protein
sequences are reportedly sufficient to change tissue and species
tropism, and viral
virulence~\cite{kuo2000retargeting,casais2003recombinant,li2006animal}.
The S protein is a trimeric transmembrane protein with protrusion on
the viral surface, like spikes, which are the key for binding and
entry to host cells.  It is composed of the receptor binding domain or
S1 subunit and S2 subunit that harbor sequences for viral fusion to
the cell
membrane~\cite{kuo2000retargeting,casais2003recombinant,li2006animal}.
Because of its importance, using the specificity of spike proteins offers an approach to classify the potential hosts of CoVs.

A common way to classify and understand the dynamics of viruses is to construct a phylogenetic tree of evolution using the sequencing data of the virus~\cite{hadfield2018a,minh_2020_iqtree2}. After the COVID-19 pandemic breakout, databases like GISAID~\cite{gisaid_website_url} collected a large number of sequence data of SARS-CoV-2 from researchers and clinicians worldwide, which provides data to conduct virus phylogenetic tree inference. Many methods have been developed and applied for constructing phylogenetic trees, including the most similar supertree algorithm (MSSA) method \cite{creevey2004does}, MRP method \cite{baum1992combining}, and approximate maximum likelihood (ML) supertree method \cite{akanni2014lu, li2020phylogenetic}, similar methods upon which the state-of-the art Nextstrain~\cite{hadfield2018a} and IQTREE2~\cite{minh_2020_iqtree2} have been developed. 
However, these methods of building a phylogenetic tree for CoVs require a high computational complexity, and the vast volume of sequences data size can cause a scalability issue for the phylogenetic-based approaches \cite{melnyk2021alpha}. For example, Nextstrain~\cite{hadfield2018a} is able to construct trees on thousands of sequences, while IQTREE2~\cite{minh_2020_iqtree2} is able to scale to tens of thousands of sequences.  There are currently millions of sequences available on  GISAID alone --- clearly viable alternatives are necessary.  Here we study machine learning clustering and classification as an alternative to phylogenetic tree building.

Some effort have been made to study the coronavirus host data~\cite{kuzmin2020machine} by using one-hot embedding (OHE) approach to get fixed length feature embedding for the spike sequence. Although OHE gives better predictive performance, one of its drawbacks is the high dimensionality of the feature vectors. Also, the columns of the OHE based vector have a linear relationship with each other, which means that one variable can be easily predicted using the other variables. This behavior can cause problems of parallelism and multicollinearity (when multiple features are correlated with each other) in high dimensions. 
Authors in~\cite{ali2021k,ali2021spike2vec} using the coronavirus spike sequences to classify different variants of COVID-19 using $k$-mers based frequency vectors. Researchers have performed clustering on the COVID-19 spike sequences using the same $k$-mers based frequency vector generation approach~\cite{ali2021effective,tayebi2021robust}. Although their approaches are effective in terms of predictive performance, the dimensionality of feature vector representation is still high, which can creates the very well known problem in machine learning the problem of the curse of dimensionality. Moreover, since for each $k$-mer, it is required to find the appropriate bin dedicated for that specific $k$-mer, this \textit{`Bin Matching'} could be expensive in terms of computational cost.
%
%

Another possible solution, which is what we propose here, is the use of the position weight matrix (PWM), sometimes also called as a position-specific weight matrix (PSWM) or position-specific scoring matrix (PSSM)~\cite{storato2021k2mem}. It is a good representation for motifs in biological sequences. A motif is a nucleotide or amino-acid sequence pattern that is widespread and has, or is conjectured to have, a biological significance. The PWM is a application of entropy and relative entropy towards identifying transcription factor binding sites (TFBSs), for example. A PWM contains information about frequency of nucleotide for each position in form of weights, these log-odds or log-probability weights are used for computing the binding affinity score. 

The PWM can be used to distinguish the binding sites from the sequence, a well known method for de novo motif sequence finding. If we do not know about a possible motif in given sequence, there are methods such as expectation maximization (EM) and Gibbs sampling which uses the PWM. Inspired by this application, we compute an absolute score from the PWM while scanning the sequence for "motifs" (here $k$-mers) of our sequence using a sliding window (of size $k$) to scan the sequence and computing absolute score as shown in Figure~\ref{fig_pwm_flow_example}. We can find relevant information of the motifs based on score calculated from the PWM. The higher the score, the more relevant the $k$-mer is.

In this paper, we propose an approach, called PWM2Vec, which is also basic implementation of position weight matrix (PWM) to generate feature vector representation from coronavirus spike sequences. Given a spike sequence, we first extract $k$-mers. From the $k$-mers, we generate the PWM (see Figure~\ref{fig_pwm_flow_diagram}). After that, we assign a score to each $k$-mer using PWM to design a feature embedding and apply machine learning methods like classification and clustering on this feature embedding. While this is inspired by methods for finding motifs (e.g., TF-promoter binding sites), here our goal to obtain a numerical representation of these $k$-mers generated from each sequence.


Our contributions in this paper are as follow:
\begin{enumerate}
    \item We propose an approach to generate fixed-length numerical representation of spike sequences using PWMs. The generated feature vectors could be used as input for any machine learning algorithm for tasks such as classification and clustering.
    \item Our proposed feature embedding approach contains more compact information and give results better than the baselines in terms of classification and clustering.
    \item Our feature vector contains fewer dimensions as compared to $k$-mers and one-hot encoding based feature vectors ($\approx 24$ times fewer dimensions than one-hot encoding and $\approx 4$ times fewer dimensions than $k$-mers based embedding), which improves the runtime for classification and clustering algorithms.
    \item We performed statistical analysis on the data and show importance of different positions of amino acids that play key role towards classification of different hosts, as well as to validate the compactness of our proposed embedding from an orthogonal point of view.
\end{enumerate}

The rest of the paper is organised as follows:
Section~\ref{sec_related_work} contains the previous work related to
the sequence classification in general and coronavirus spike sequence
classification in particular. Section~\ref{sec_proposed_approach}
contains the detail about our proposed alignment free method for spike
sequence classification. Section~\ref{sex_experimental_setup} contains
the experimental setup and dataset collection and statistics detail. The results for
our proposed method are given in
Section~\ref{sec_results_discussion}. Finally, we conclude our paper
in Section~\ref{sec_conclusion}.

\section{Related Work}\label{sec_related_work}

Several machine learning approaches based on $k$-mers have been proposed in the literature for classification and clustering tasks~\cite{solis-2018-hiv,queyrel-2020-metagenomic,ali2021k,ali2021effective}. More specifically, there are a lot of classical algorithms for sequence classification~\cite{wood-2014-kraken,kawulok-2015-cometa}. Although these methods are proven to be useful in respective studies, it is not clear if they can be used in context of coronavirus data. Also, another major problem with all those methods is high computational complexity of the algorithms (because of high dimensional representation of data), which can result in the higher runtime of the underlying classification algorithms.

Postition weight matrix (PMW) based approaches have been successfully applied for diverse sequence analysis, motifs predictions and identification studies. Several popular software applications or web servers have been build based on the implementation of PWM, e.g., the PWMscan software package \cite{ambrosini2018pwmscan} and PSI-BLAST \cite{bhagwat2007psi}. Also, a many other advanced algorithms have been implemented to optimize PWM technique: examples include MEME (multiple EM for motif elicitation) \cite{bailey1994fitting} implemented based expectation maximum (EM) and Gibbs Sampler \cite{lawrence1993detecting} for de novo motifs discovery that uses Gibbs sampling algorithms \cite{xia2012position}, \cite{hashim2019review}. The MEME EM algorithm basically finds an initial motif and repeatedly use EM steps improve motifs until the PWM values do not improve further \cite{sinha2003ymf, bailey1994fitting}.  Also, the BaMM (Bayesian Markov model) algorithm build based on the Markov to model correlations among nucleotides at other positions --- since the PWM cannot because the method assumes probabilities at different sites are independent from each others~\cite{hashim2019review}. The PWM method continues to be applied and extended. Log2PWMs is simple implementation of PWM extended to enable conversion or reconstruction of PWM representation from sequence logo \cite{gao2017logo2pwm}. 

The PWM is also used for the binding specificity of a transcription factor (TF)~\cite{yang2015dna}. It can be used to scan a sequence for the presence of DNA words, which are comparatively more similar to the PWM than to the background~\cite{stormo2000dna,boeva2016analysis}.
Authors in~\cite{wright2011occupancy} evaluate the Bayesian network and support vector machine (SVM) algorithm on four distinct TF binding sites based data sets and analyze their performance using PWM. Authors in \cite{bi2011tree} develop a tree-based PWM algorithm to accurately simulate the interaction between TF and its binding sites. A new di-nucleotide PWM approach is proposed in~\cite{nandi2012optimizing}, which outperforms the conventional mono-nucleotide PWM method. Moreover, the research done in~\cite{wu2011improved} proposes an improved position weight matrix (IPWM) method to recognize the prokaryotic promoters based on an entropy measurement. Using the Hepatitis C Virus (HCV) nucleotide sequences, authors in~\cite{qiu2009hcv} designed a global PWM for the genotype of HCV genomes. Then using the PWM, signatures were selected from the 5' NCR, CORE, E1, and NS5B regions of the HCV genome. The predictive power of the selected signatures is then evaluated for predicting the most common HCV genotypes and subtypes. 

Aside from DNA analysis, the PWM can also be applied to amino acid sequences. Authors in \cite{wang2017predicting} develop an approach involving position specific scoring matrix (PSSM), another name of PWM, to predict protein-protein interactions between protein sequences. They first transform each protein into a PSSM, and then adopt the PSSMs to detect distantly related proteins as well as the protein quaternary structural attributes and protein folding patterns. The research of in~\cite{hiller2004predisi} proposes a PWM-based algorithm to predict signal peptide sequences and their cleavage positions in the amino acid sequences of bacteria and eukaryotes. Authors in \cite{cheol2010position} develop a PWM-based method for protein function prediction and propose an argument for why the PWM and associated features have great potential for protein sequence analysis.
Although the above methods are successful in respective domains, they do not propose a general method to design a feature embedding for the underlying sequence, which contains rich information about the sequence and can be used as input to various machine learning algorithm.

Designing efficient feature vector based representations haave been studied in many domains such as graph analytics~\cite{ali2021predicting,AHMAD2020Combinatorial}, smart grid~\cite{ali2019short,ali2019short_AMI}, electromyography (EMG)~\cite{ullah2020effect}, clinical data analysis~\cite{ali2021efficient}, network security~\cite{ali2020detecting}, and text classification~\cite{Shakeel2020LanguageIndependent}. After the spread of COVID-19, efforts have been made to study the behavior of the virus using machine learning approaches. Several methods have been proposed recently for the classification of spike sequences. 
Authors in~\cite{ali2021k,ali2021classifying} uses $k$-mers along with kernel based approach to classify the spike sequences. Authors in~\cite{kuzmin2020machine} propose the use of one-hot encoding to classify the viral hosts of coronavirus using spike sequences only. Although they were able to achieve higher predictive performance, authors in~\cite{ali2021k} shows that $k$-mers based approach performs better than the one-hot since it preserves sequence order information more effectively. An efficient clustering of the spike sequences is done in~\cite{ali2021effective}.

\section{Proposed Approach}
\label{sec_proposed_approach}

In this section, we propose an approach, PWM2Vec, to generate a fixed-length numerical feature embedding from coronavirus spike sequences for the purposes of host-specification. We also discuss the baseline approaches, specifically one-hot embedding (OHE)~\cite{kuzmin2020machine,ali2021spike2vec} and $k$-mers based feature embedding~\cite{ali2021k,ali2021spike2vec}. 
We perform feature selection using ridge regression~\cite{hoerl1975ridge} on the resulting embedding before applying machine learning (ML) algorithms.  This helps to reduce the dimensionality of the embedding, hence the training time of the downstream ML algorithms.

\subsection{One-Hot Embedding (OHE)}
Machine learning algorithms requires the input to be in a numerical format, it is necessary to processes the (spike protein) sequence data into some numerical representation to apply these algorithms. One-hot embedding (OHE)~\cite{kuzmin2020machine} is a typical approach for obtaining a fixed-length numerical representation from sequence data.  Considering an alphabet $\Sigma$, which contains the characters (amino acids) of the spike protein sequence, we need to map each character of $\Sigma$ to a numerical (binary 0-1 vector) representation. We have 24 unique amino acids in the protein sequence data, namely,
"ABCDEFGHIJKLMNPQRSTVWXYZ". For each character in the protein sequence, we design a feature vector wherein we encode each symbol separately.  Each symbol has length $24$ and has a value of 1 at the location corresponding to the position of the character in the alphabet, and 0 at all other places in the vector. For example, amino acid Cysteine (C) is encoded as $001\dots0$. We then concatenate the numerical representations all characters of the protein sequence into a single binary feature vector of this spike sequence. In our coronavirus spike protein sequence dataset, after multiple alignment, the length of each spike sequence is $3498$. Therefore, the length of each binary vector computed using OHE would be $3498 \times 24 = 83,952$.

\subsection{$k$-mers based Frequency Vectors}
\begin{figure}[!ht]
  \centering
  \includegraphics[scale = 0.4]{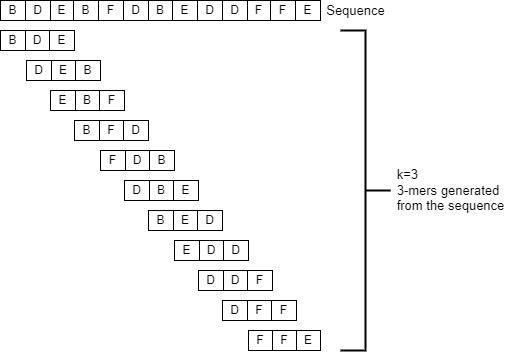}
  \caption{Generating 3-mers ($k=3$) from a spike protein sequence using a sliding window. Since this sequence has length 13, the total number of $k$-mers generated is 13 - 3 +1 = 11 (Equation~\ref{eq:totalkmers}).}
  \label{fig_kmer_flow}
\end{figure}

One of the major drawbacks of OHE is the high dimensionality of the resulting set of feature vectors. Another problem with OHE is that some (sequential) ordering information on the characters (amino acids) of the sequence is not preserved. An approach which addresses both of these problems is to use sub-strings (also called mers) of length $k$, i.e., $k$-mers. From a sequence, $k$-mers are generated by applying a sliding window of size $k$ over the sequences (see Figure~\ref{fig_kmer_flow}).
Given a sequence of length $N$, the total number of k-mers that could be generated is as follows:
\begin{equation}
    \text{Total $k$-mers = }(N - k) + 1~.
\label{eq:totalkmers}
\end{equation}

\begin{remark}
    In our experiments to generate k-mers based frequency vectors, we use $k=3$ (as done in~\cite{ali2021k,ali2021spike2vec}). 
\end{remark}

After generating the $k$-mers, we create a feature vector $\Phi$ (a frequency vector), which contains the frequency (count) of each $k$-mer occurring in the sequence~\cite{ali2021k,ali2021spike2vec}. Given some sequence $\sigma$ on alphabets $\Sigma$, the length of feature vector $\Phi_k (\sigma)$ will be $\vert \Sigma \vert^k$. Since we are working with spike protein (amino acid) sequences, and taking $k=3$ in our experiments, the length of the feature vector we use is $24^3 = 13824$. This feature vector can be used as input for various ML algorithms. Note that generating sub-strings with a sliding window preserves some (sequential) ordering information on the characters (amino acids) of the (spike protein) sequences, which counters one of the drawbacks of the OHE approach. But still to get the frequency count for k-mers we got a problem of high computational cost for bin matching, especially for worst case searches. Also the dimensionality of the frequency vectors is still high.

\subsection{PWM2Vec}
Despite the fact that the problem of high dimensionality of the vectors generated in the OHE approach is somewhat mitigated in the $k$-mers approach, the dimensionality of the frequency vectors generated in the $k$-mers approach remains quite high --- further improvements can certainly be made. Also, if we can reduce the computational cost for bin matching that would be huge improvement in computational cost. To address these problems, we propose PWM2Vec, an approach for generating a fixed-length numerical feature vector based on the concept of the position-weight matrix~\cite{stormo-1982-pwm}.

While our approach is inspired by the value of the PWM for finding motifs in (e.g., protein) sequences, we use it in a slightly different way here.  Here, we build a PWM from the $k$-mers of our sequence, and our feature vector is the score of each $k$-mer from this PWM.  This allows us to take advantage of $k$-mers in their ability to capture locality information, while also capturing the importance of the position of each amino acid in the sequence (information that is lost in computing $k$-mer frequency vector).  Combining these pieces of information in this way allows us to devise a compact and general feature embedding that can be used with many downstream ML tasks.

\begin{figure}[!h]
  \centering
  \includegraphics[scale = 0.365] {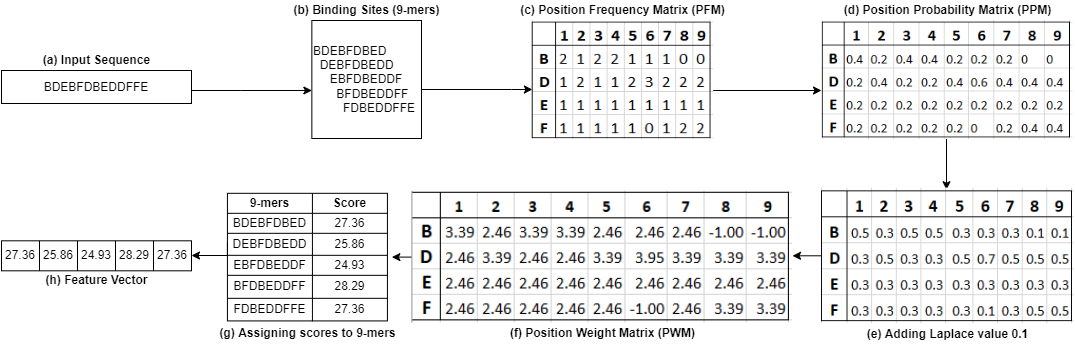}
  \caption{Building feature vector for the classification models by computing the Position Weight Matrix from the k-mers of a sequence.}
  \label{fig_pwm_flow_diagram}
\end{figure}

Our approach PWM2Vec for feature vector generation, is summarized in Figure~\ref{fig_pwm_flow_diagram}, we follow the steps (a--h) as explained below. Figure~\ref{fig_pwm_flow_diagram} (a) Given an input spike protein sequence , Figure~\ref{fig_pwm_flow_diagram} (b) we first extract the $k$-mers (we use $k = 9$ in the experiments, which is decided using a standard validation set approach~\cite{validationSetApproach}). 
Figure~\ref{fig_pwm_flow_diagram} (c) We then generate a position frequency matrix , which contains the frequency count for each character at each position. Note that, in the example, since the (amino acid) sequence is composed of four characters, there are four possible characters at any position.  At position 1, for example, in all 5 $k$-mers, there are 2 B characters, and so the frequency count of B at position 1 is 2. In our experiments, since we have $24$ unique amino acids in our spike protein sequence dataset, our PFMs will have 24 rows and $k=9$ columns. Figure~\ref{fig_pwm_flow_diagram} (d) Next, we normalize the PFM matrix, and create a position probability matrix (PPM) containing the probability of each (amino acid) character at each position. For example, the probability of B in the $k$-mers at position 1 is the following:
\begin{equation}
    \frac{\text{frequency count}}{\text{total count}} = 2/5 = 0.4
\end{equation}

It is possible that the frequency (hence probability in the PPM) of a character at a certain position is 0. In order to avoid 0 values at any position in the matrix while calculating the probability, we add a Laplace estimator, or a pseudocount to each value in the position probability matrix as shown in Figure~\ref{fig_pwm_flow_diagram} (e). We use a pseudocount of 0.1 in our experiments~\cite{nishida2009pseudocounts}. We then compute a position weight matrix (PWM) from the adjusted probability matrix. We make the PWM by computing the log likelihood of each amino acid character $c$, i.e., $c \in A,B,C,\dots,Z$, appearing at each position $i$ according to
\begin{equation}
    W_{c, i}=\log_{2} \frac{p(c, i)}{p(c)}  \text{ \{where } c \in A, B, C...Z(bases) \}
\end{equation}
Note that this likelihood is taken under the assumption that the expected frequency of each amino acid is the same (i.e., $p(c) = 1/|\Sigma|$).
since we have 24 bases amino acids($p(c) = 1/24 = 0.041$). Figure~\ref{fig_pwm_flow_diagram} (f) shows the computed position weight matrix (PWM).

\begin{figure}[!h]
  \centering
\includegraphics[scale = 0.55]{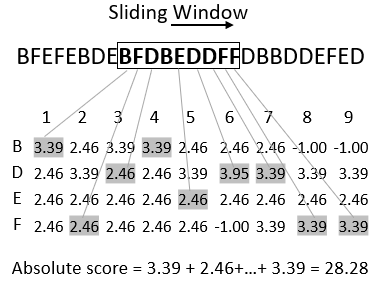}
  \caption{Computing the Score from PWM using sliding window on a sequence for a 9-mer.}
  \label{fig_pwm_flow_example}
\end{figure}

After getting the PWM, we use it to compute the absolute scores for each individual k-mer generated from the sequence  (see Figure~\ref{fig_pwm_flow_diagram} (g) for an example). It is sum of score of bases for the index. Calculating the absolute score is shown in Figure~\ref{fig_pwm_flow_example} for k-mer (BFDBEDDFF). The highlighted values in matrices are summed up to give Absolute score for the k-mer which sums up to $28.28$.


Finally, the scores of all the 9-mers are concatenated to get the final feature vector for the given sequence (see Figure~\ref{fig_pwm_flow_diagram} (h) for an example). This whole process is repeated for each sequence.

Given a k-mer and PWM, the score for that k-mer can be computed as given in Figure~\ref{fig_pwm_flow_example}. The final length of PWM2Vec based feature vector is $3490$, which is equal to the number of k-mers in each spike sequence.

\subsection{Feature Selection Method}
\begin{figure}[!ht]
  \centering
  \includegraphics[scale = 0.5]{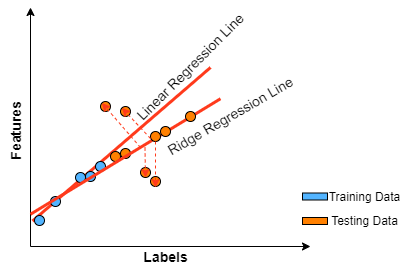}
  \caption{Ridge Regression diagram. With alpha value tending to a large positive number, the slope will tend to zero thereby reducing variance in the dataset.}
  \label{fig_ridge_regression}
\end{figure}

We use Ridge Regression (RR) as the feature selection approach, which is a commonly used for estimating parameters thereby addressing collinearity in multiple linear regression model problems~\cite{mcdonald2009ridge,ali2021simpler}. This method uses a bias to boost the performance of the model by improving the variance and making the slope more horizontal like.  This is useful when we need to find out which of the independent attributes are not needed. This gives us the option of removing such columns (attributes) and bring the slope to zero (see Figure~\ref{fig_ridge_regression}). The expression for performing ridge regression is the following:
\begin{equation}
    min(\text{Sum of square residuals} + \alpha \times \text{slope}^2)
\end{equation} 
where $\alpha \times {slope}^2$ is a penalty term. 
The total number of selected attributes after performing RR in OHE is $22322$, for $k$-mers approach is $7088$  while $1616$ features are selected for the PWM2Vec based approach.

\section{Experimental Setup}\label{sex_experimental_setup}
In this section, we describe the setup we used for the experiments followed by the dataset statistics. We also give visual representation of the data using t-SNE plots.
All experiments are conducted using an Intel(R) Xeon(R) CPU E7-4850 v4
@ $2.40$GHz having Windows 10 $64$ bit OS with 32 GB memory.  We Implemented our algorithm in Python
and the code is available online for
reproducibility\footnote{\url{https://github.com/sarwanpasha/PWM2Vec}}. Our
pre-processed data is also available
online\footnote{Available in published version}. For classification, we use Support Vector Machine (SVM), Naive Bayes (NB), Multiple Linear Regression (MLP), K Nearest Neighbors (KNN), Random Forest (RF), Logistic Regression (LR), and Decision Tree (DT) with default parameters. 

\subsection{Evaluation Metrics}
To measure the performance of the classifiers, we use average accuracy, precision, recall, weighted and macro F1, and receiver operating characteristic curve ``ROC" area under the curve ``AUC" (one-vs-rest approach). We also measure the training time (in seconds) for each of the classifier.

For clustering, we use simple k-means algorithm and use $3$ internal clustering quality metrics namely Silhouette Coefficient, Calinski-Harabasz Score, and Davies Bouldin Score to measure the performance of the clusters. We also show the runtime for k-means for different embedding methods.

\subsubsection{Silhouette Coefficient}
The silhouette coefficient~\cite{rousseeuw1987silhouettes} is used for interpretation and validation of consistency within clusters of data. A clustering algorithm having well-defined (comparatively pure) clusters will have a higher silhouette coefficient value. The Silhouette Coefficient (SC) is computed as follows:

\begin{equation}
    {SC = \frac{y - x}{max\{y,x\}}}
\end{equation}
where $x$ is the average distance between a sample and all other points in the data belonging to the same class and $y$ is the average distance between sample $x$ and all other data points in the next nearest cluster.

\subsubsection{Calinski-Harabasz Score}
The Calinski-Harabasz score~\cite{calinski1974dendrite} is used to measure the quality of a clustering algorithm based on the mean between and inside cluster sum of squares. A clustering algorithm with well defined clusters will have a higher Calinski-Harabasz score. The Calinski-Harabasz score is defined as the ratio of the between-clusters dispersion (the sum of distances squared) mean and the within-clusters dispersion. More formally, given a dataset $D$ of size $n_D$ that has been clustered into $j$ clusters, we use following expression to compute Calinski-Harabasz (CH) score:

\begin{equation} 
    CH = \frac{tr(B_j)}{tr(W_j)} \times \frac{(n_D-j)}{(j-1)}
\end{equation}
where $tr(B_j)$ is the trace of the between cluster dispersion matrix, $tr(W_j)$ is the trace of the within-group dispersion matrix.

\subsubsection{Davies-Bouldin Score}
The Davies-Bouldin (DB) score~\cite{davies1979cluster} of a clustering $C$ is defined as follows:
\begin{equation} 
  DB(C)= \frac{1}{|C|} \sum_{i=1}^{|C|} max_{j \leq |C|,j \neq i}
  D_{ij}
\end{equation}
where $D_{ij}$ is the ratio of the ``within-to-between cluster
distance'' of the $i$th and $j$th clusters. For each cluster, we compute the worst case ratio ($D_{ij}$) of a within-to-between cluster distance between it and any
other cluster, and then take the mean. Therefore, by minimizing the DB score, we can make sure that different clusters are separate from each other (smaller value is better).


\subsection{Dataset  Collection and Statistics}
\begin{figure}[!ht]
  \centering
  \includegraphics[scale=0.47]{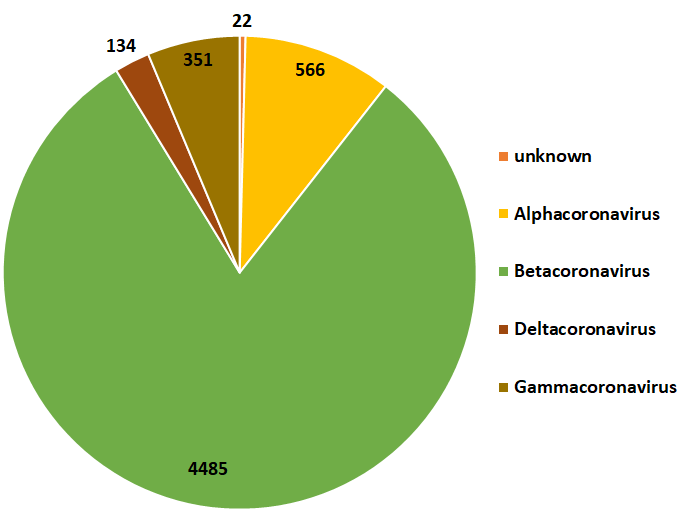}
  \caption{Pie chart for the distribution of different Genus in the dataset.}
  \label{fig_genus}
\end{figure}

\begin{figure}[!ht]
  \centering
  \includegraphics[scale=0.36]{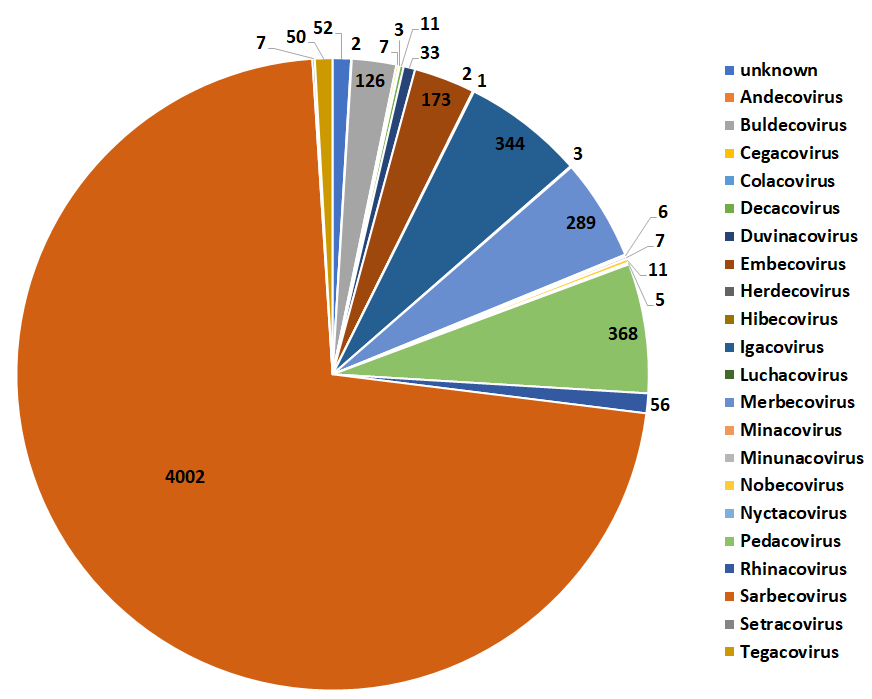}
  \caption{Pie chart for the distribution of different SubGenus in the dataset.}
  \label{fig_sub_genus}
\end{figure}

The Spike protein sequences of CoVs for all hosts used in this analysis were retrieved (on September 21, 2021) from the NIAD Virus Pathogen Database and Analysis Resource (ViPR)~\cite{pickett2012vipr} and GISAID~\cite{gisaid_website_url}. A total of $5568$ complete protein sequence were collected (3358 from ViPR and 2210 from GISAID), later dropping $10$ not attributable to any host detail. The distribution of the dataset across the different host types (grouped by family) is shown in Table~\ref{tbl_dataset_statistics}, which contains information about the $21$ host types that we collected from the annotation of the sequences.
We also divided the viral sequences themselves into genus and subgenus to see which category a specific coronavirus belongs to. Figure~\ref{fig_genus} and Figure~\ref{fig_sub_genus} contains distributions of viral genus and subgenus, respectively.
The multiple sequence alignment (MSA) for the sequence dataset was conducted using Mafft Alignment software with default parameter settings which automatically select the appropriate strategy according to the sequence data size.  In our case, the gap opening penalty \texttt{op} was 1.53 and gap extension penalty \texttt{ep} was 0.123~\cite{rozewicki2019mafft}.  Given that our dataset was already sufficiently large, and contained a number of unknown or identified amino acids, we were constrained to using the minimum accuracy parameter of Mafft MSA in order to allow the alignment to complete in a reasonable amount of time.  Attempts to set more stringent parameters (\texttt{op} and \texttt{ep}) in order to improve this alignment resulted in runtimes $>24$ hours.  Since this is already the case when performing multiple alignment of even ~5000 sequences, anything substantially larger is out of reach. 

\begin{table}[h!]
    \centering
    \begin{tabular}{cc|cc}
    \hline
        Host Name & \# of Sequences & Host Name & \# of Sequences \\
        \hline \hline
        Humans & 1813 & Rats & 26 \\
        Environment &  1034 & Pangolins & 21 \\
        Weasel & 994 & Hedgehog & 15 \\
        Swines & 558 & Dolphin & 7 \\
        Birds & 374 & Equine & 5 \\
        Camels & 297 & Fish & 2 \\
        Bats & 153 & Unknown & 2 \\
        Cats & 123 & Python & 2 \\
        Bovines & 88 & Monkey & 2 \\
        Canis & 40 & Cattle & 1 \\
        Turtle & 1 && \\
        \hline
    \end{tabular}
    \caption{Dataset Statistics for 5558 coronavirus hosts.}
    \label{tbl_dataset_statistics}
\end{table}

\subsection{Data Visualization}
\begin{figure}[h!]
    \begin{minipage}{.5\textwidth}
        \centering
          \includegraphics[scale=0.25]{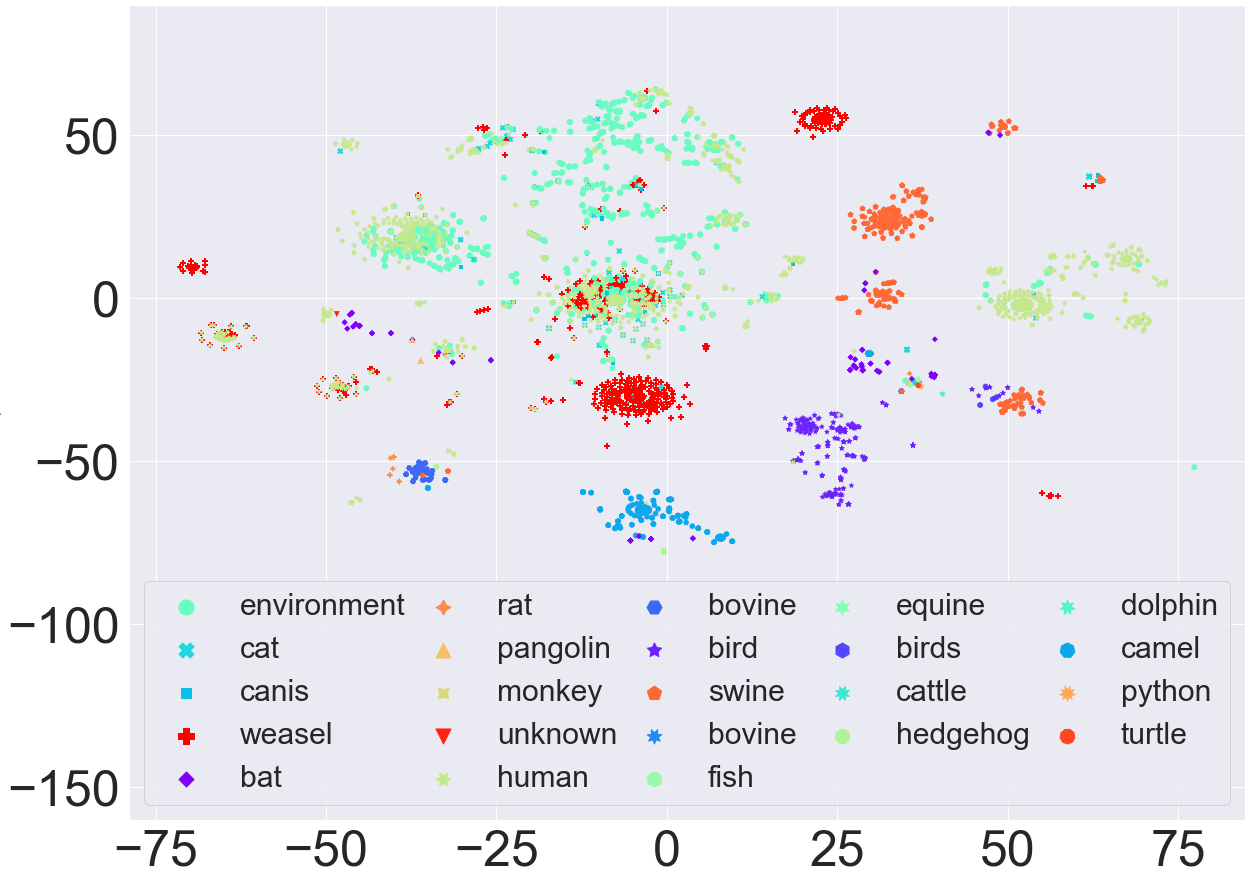}
          \caption{t-SNE plots for OHE}
          \label{fig_tsne_ohe}
        \end{minipage}%
        \begin{minipage}{.5\textwidth}
        \centering
          \includegraphics[scale=0.25]{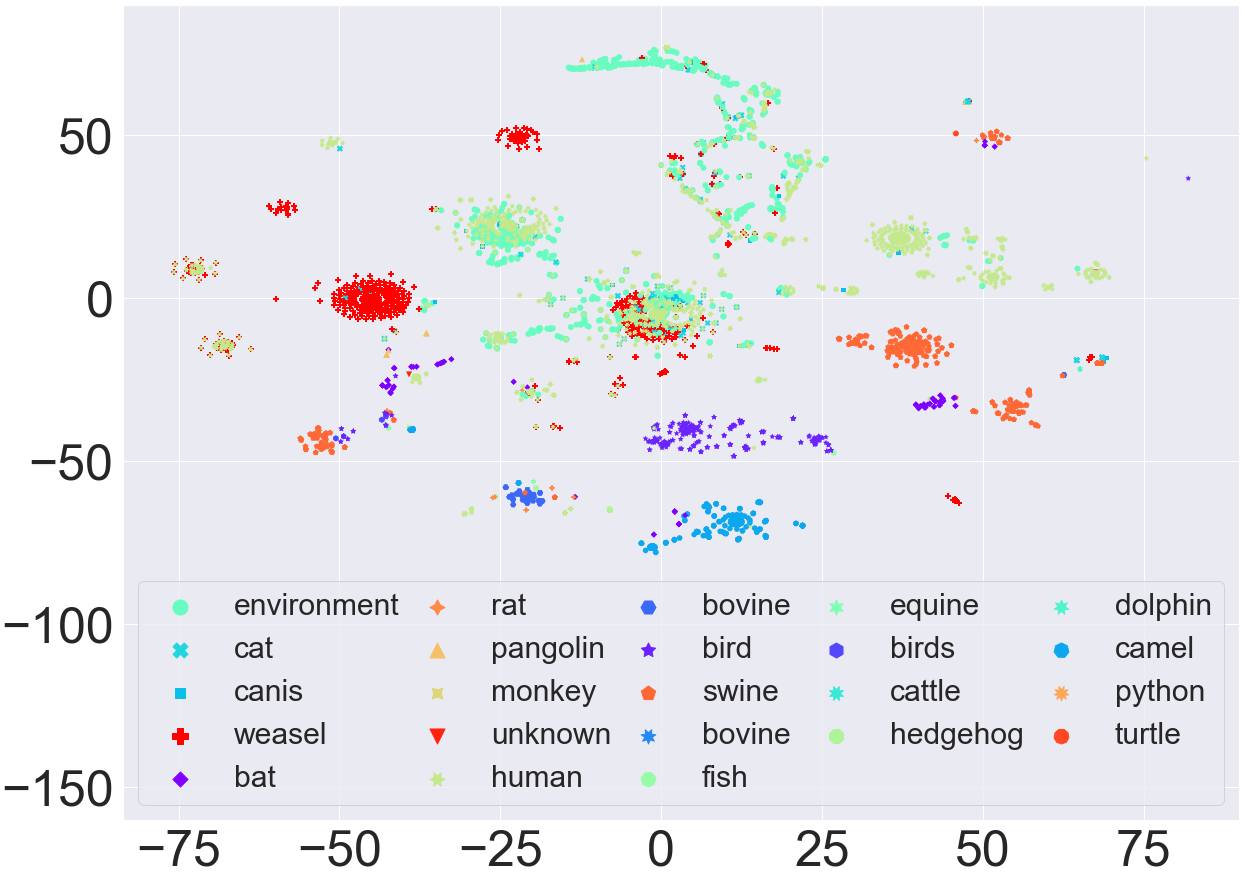}
            \caption{t-SNE plots for k-mers}
            \label{fig_tsne_kmers}
    \end{minipage}
\end{figure}

\begin{figure}[h!]
  \centering
  \includegraphics[scale=0.38]{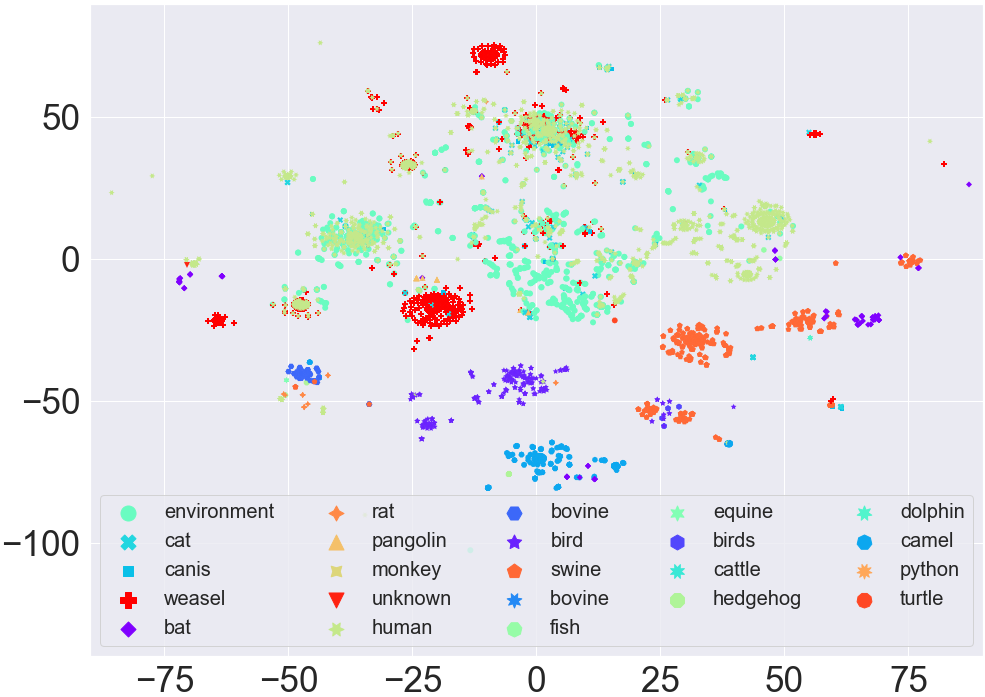}
  \caption{t-SNE plots for PWM2Vec}
\label{fig_tsne_pwm}
\end{figure}

In order to see if there is any natural (hidden) clustering in the data, we used t-distributed stochastic neighbor embedding (t-SNE)
\cite{van2008visualizing} approach, which maps input sequences to 2D real
vectors. 
The t-SNE plots for different embedding methods are shown in Figure~\ref{fig_tsne_ohe}, Figure~\ref{fig_tsne_kmers}, and Figure~\ref{fig_tsne_pwm} contains the t-SNE plots for OHE, $k$-mers, and PWM2Vec, respectively. We can observe that although with PWM2Vec, more information is included in less dimensional feature vectors, the proposed embedding approach was able to preserve the structure of data similar to OHE and k-mers.

\section{Results and Discussion}\label{sec_results_discussion}
In this section, we present our results for PWM2Vec and compare its performance with the baseline One-Hot Embedding (OHE) and the more recent $k$-mers based embedding approach which has shown to be an improvement over OHE~\cite{ali2021k,ali2021spike2vec}.  For classification, we also show the results for feature selection method (ridge regression) for all embedding approaches. We also show the effect of runtime with increasing number of sequences for the best performing classification algorithm. In the end, we show some statistical analysis on the data as well as on the feature vectors computed using different approaches.

\subsection{Classification Results}
Table~\ref{tbl_no_feat_selec_results} shows the results for different embedding methods with different classification methods without performing any feature selection on the feature vectors. We can see that RF with PWM2Vec is consistently performing better than other embedding methods (in some cases the performance gain for PWM2Vec on third significant digit). Also in terms of training runtime, the NB classifier with PWM2Vec is much better than other approaches. This behavior shows that PWM2Vec is not only better in terms of predictive performance, but is also better in terms of runtime.

Table~\ref{tbl_ridge_feat_selec_results} contains the classification results after applying ridge regression classifier of different feature embedding approaches. We can again see that RF with PWM2Vec outperforms other embedding methods for majority of the metrics and NB with PWM2Vec is best in terms of runtime.

\begin{table*}[h!]
    \centering
    \begin{tabular}{p{1.5cm}p{0.6cm}p{0.5cm}p{0.5cm}p{0.7cm}p{0.9cm}p{0.9cm}p{0.7cm}|p{1.1cm}}
    \hline
        & & Acc. & Prec. & Recall & F1 (Weig.) & F1 (Macro) & ROC AUC & Train Time (Sec.) \\
        \hline \hline
        
        \multirow{7}{1.5cm}{OHE}
        & SVM & 0.82 & 0.83 & 0.82 & 0.82 & 0.70 & 0.83 & 389.128 \\
        & NB & 0.67 & 0.80 & 0.67 & 0.65 & 0.64 & 0.81 & 56.741 \\
        & MLP & 0.77 & 0.76 & 0.77 & 0.75 & 0.44 & 0.71 & 390.289 \\
        & KNN & 0.80 & 0.79 & 0.80 & 0.79 & 0.55 & 0.78 & 16.211 \\
        & RF & 0.83 & 0.83 & 0.83 & 0.82 & 0.66 & 0.83 & 151.911 \\
        & LR & 0.83 & 0.84 & 0.83 & 0.82 & 0.71 & 0.83 & 48.786 \\
        & DT & 0.82 & 0.83 & 0.82 & 0.81 & 0.64 & 0.81 & 21.581 \\

        \hline
        \multirow{7}{1.2cm}{k-mers}
         & SVM & 0.81 & 0.82 & 0.81 & 0.81 & 0.63 & 0.83 & 52.384 \\
         & NB & 0.65 & 0.77 & 0.65 & 0.64 & 0.46 & 0.74 & 9.031 \\
         & MLP & 0.81 & 0.82 & 0.81 & 0.81 & 0.52 & 0.77 & 44.982 \\
         & KNN & 0.80 & 0.80 & 0.80 & 0.79 & 0.54 & 0.75 & 2.917 \\
         & RF & 0.83 & 0.84 & 0.83 & 0.82 & 0.63 & 0.82 & 17.252 \\
         & LR & 0.82 & 0.84 & 0.82 & 0.82 & 0.68 & 0.83 & 48.826 \\
         & DT & 0.81 & 0.82 & 0.81 & 0.81 & 0.62 & 0.81 & 4.096 \\
        \hline
        \multirow{7}{1.2cm}{PWM2Vec}
        & SVM & 0.80 & 0.81 & 0.80 & 0.81 & 0.71 & \textbf{0.85} & 40.55 \\
        & NB & 0.46 & 0.69 & 0.46 & 0.40 & 0.47 & 0.76 & \textbf{1.56} \\
        & MLP & 0.80 & 0.81 & 0.80 & 0.79 & 0.57 & 0.78 & 17.28 \\
        & KNN & 0.82 & 0.81 & 0.82 & 0.81 & 0.58 & 0.79 & 2.86 \\
        & RF & \textbf{0.85} & \textbf{0.85} & \textbf{0.85} & \textbf{0.84} & \textbf{0.71} & 0.84 & 5.44 \\
        & LR & 0.82 & 0.82 & 0.82 & 0.81 & 0.71 & 0.84 & 43.35 \\
        & DT & 0.81 & 0.81 & 0.81 & 0.81 & 0.64 & 0.83 & 3.46 \\

        \hline
    \end{tabular}
    \caption{Performance comparing for different embedding methods and different classifiers without using any feature selection approach. Best values are shown in bold.}
    \label{tbl_no_feat_selec_results}
\end{table*}

\begin{table*}[h!]
    \centering
    \begin{tabular}{p{1.5cm}cp{0.5cm}p{0.5cm}p{0.7cm}p{1.1cm}p{1.1cm}p{0.7cm}|p{1.1cm}}
    \hline
        & & Acc. & Prec. & Recall & F1 (Weig.) & F1 (Macro) & ROC AUC & Train Time (Sec.) \\
        \hline \hline
        
        \multirow{7}{1.5cm}{OHE}
         & SVM & 0.83 & 0.83 & 0.83 & 0.82 & 0.67 & 0.81 & 63.992 \\
         & NB & 0.63 & 0.75 & 0.63 & 0.61 & 0.53 & 0.77 & 9.436 \\
         & MLP & 0.82 & 0.82 & 0.82 & 0.80 & 0.51 & 0.75 & 64.636 \\
         & KNN & 0.78 & 0.78 & 0.78 & 0.78 & 0.61 & 0.81 & 2.730 \\
         & RF & 0.83 & 0.83 & 0.83 & 0.82 & 0.59 & 0.81 & 22.423 \\
         & LR & 0.83 & 0.82 & 0.83 & 0.83 & 0.65 & 0.84 & 26.094 \\
         & DT & 0.83 & 0.83 & 0.83 & 0.82 & 0.60 & 0.83 & 6.316 \\
        \hline
        \multirow{7}{1.2cm}{k-mers}
         & SVM & 0.81 & 0.81 & 0.81 & 0.81 & 0.73 & 0.87 & 30.877 \\
         & NB & 0.67 & 0.78 & 0.67 & 0.67 & 0.64 & 0.83 & 4.012 \\
         & MLP & 0.83 & 0.83 & 0.83 & 0.83 & 0.58 & 0.81 & 26.280 \\
         & KNN & 0.80 & 0.80 & 0.80 & 0.80 & 0.69 & 0.83 & 1.601 \\
         & RF & 0.82 & 0.82 & 0.81 & 0.82 & 0.73 & 0.87 & 6.786 \\
         & LR & 0.82 & 0.82 & 0.81 & 0.82 & 0.79 & 0.88 & 39.501 \\
         & DT & 0.83 & 0.84 & 0.83 & 0.83 & 0.70 & 0.88 & 2.429 \\
         \hline
        \multirow{7}{1.2cm}{PWM2Vec}
        & SVM & 0.78 & 0.79 & 0.78 & 0.78 & 0.75 & 0.89 & 24.53 \\
        & NB & 0.41 & 0.64 & 0.41 & 0.40 & 0.38 & 0.68 & 0.94 \\
        & MLP & 0.81 & 0.81 & 0.81 & 0.80 & 0.67 & 0.82 & 9.85 \\
        & KNN & 0.80 & 0.80 & 0.80 & 0.79 & 0.62 & 0.80 & 1.55 \\
        & RF & \textbf{0.84} & \textbf{0.84} & \textbf{0.84} & \textbf{0.85} & \textbf{0.80} & 0.86 & 5.06 \\
        & LR & 0.80 & 0.81 & 0.80 & 0.80 & 0.66 & \textbf{0.90} & 21.76 \\
        & DT & 0.80 & 0.80 & 0.80 & 0.80 & 0.64 & 0.82 & 2.00 \\

        \hline
    \end{tabular}
    \caption{Performance comparing for different embedding methods and different classifiers using ridge regression as the feature selection approach.  Best values are shown in bold.}
    \label{tbl_ridge_feat_selec_results}
\end{table*}

\subsubsection{Effect of Runtime}
\begin{figure}[h!]
    \begin{minipage}{.5\textwidth}
        \centering
        \includegraphics[scale = 1] {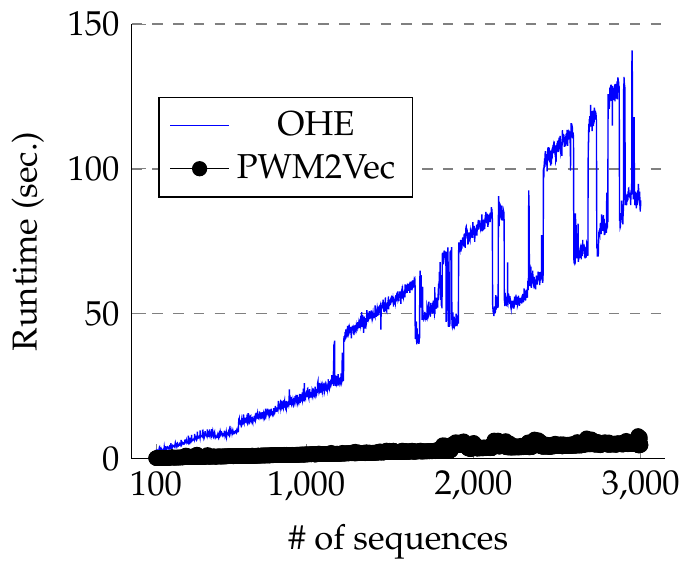}
        \end{minipage}%
        \begin{minipage}{.5\textwidth}
        \centering
        \includegraphics[scale = 1]{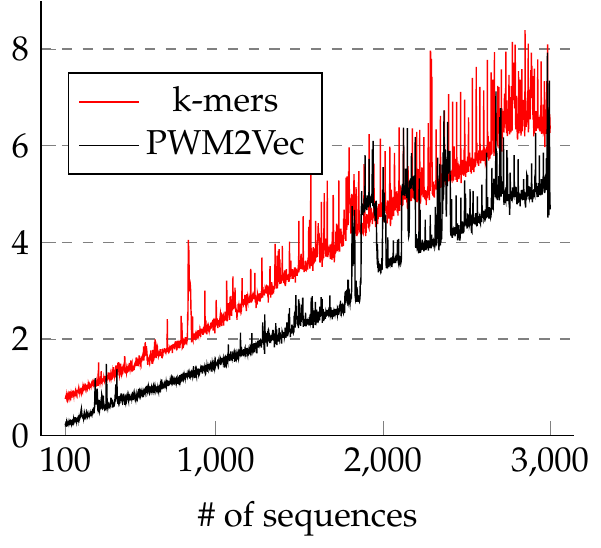}
    \end{minipage}
    \caption{Runtime comparison for different embedding methods with increasing number of sequences using random forest classifier (best performing classifier). The figure is best seen in color.}
    \label{fig_runtime_effect_classification}
\end{figure}

To evaluate the effect of runtime on the sequences, we use the best performing classifier (Random Forest) and use different embedding methods to perform classification with increasing number of sequences. Figure~\ref{fig_runtime_effect_classification} shows the effect of runtime for (a) OHE vs PWM2Vec and (b) k-mers vs PWM2Vec. We can see that in both cases, PWM2Vec is better in terms of runtime as we increase the number of sequences (on x-axis).

\subsection{Clustering Results}
\begin{figure}[h!]
  \centering
  \includegraphics[scale=0.5]{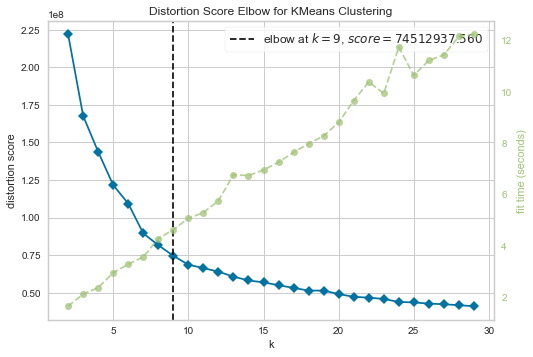}
  \caption{Elbow method for optimal number of clusters}
  \label{fig_elbow_method}
\end{figure}

For clustering, we use the same feature embeddings that we use for classification task. To get the optimal number of clusters, we used the Elbow method~\cite{ali2021effective}. This method for different number of clusters (ranging from 2 to 30) is performing clustering to see the trade-off between the runtime and the sum of squared error (distortion score). The optimal number of clusters selected are $9$ (see Figure~\ref{fig_elbow_method}). 

For the purposes of clustering, we use the simple k-means algorithm. The results of clustering on different embedding methods are shown in Table~\ref{tbl_cluster_results_new}. We can see that PWM2Vec is better in terms of Silhouette Coefficient and runtime while k-mers based approach is better in terms of Calinski-Harabasz Score and Davies-Bouldin Score.

\begin{table}[ht!]
  \centering
  \begin{tabular}{p{1.9cm}p{1.7cm}p{2.4cm}p{2.4cm}p{1.5cm}}
    \hline
    & \multicolumn{3}{c}{Evaluation Metrics} \\
    \cline{2-4}
    Methods & Silhouette Coefficient & Calinski-Harabasz Score & Davies-Bouldin Score & Runtime (Sec.) \\
    \hline	\hline	
    OHE & 0.631 & 2210.343 & 1.354 & 177.54 \\
    k-mers & 0.735 & \textbf{14296.17} & \textbf{0.534} & 36.57 \\
    PWM2Vec & \textbf{0.750} & 2563.547 & 1.314 & \textbf{23.67} \\
    \hline
  \end{tabular}
  \caption{Internal Clustering quality metrics for k-means. Best values are show in bold.}
  \label{tbl_cluster_results_new}
\end{table}




\subsection{Statistical Analysis on Data}
\begin{figure}[ht!]
  \centering
  \centering
  \includegraphics[scale = 0.9]{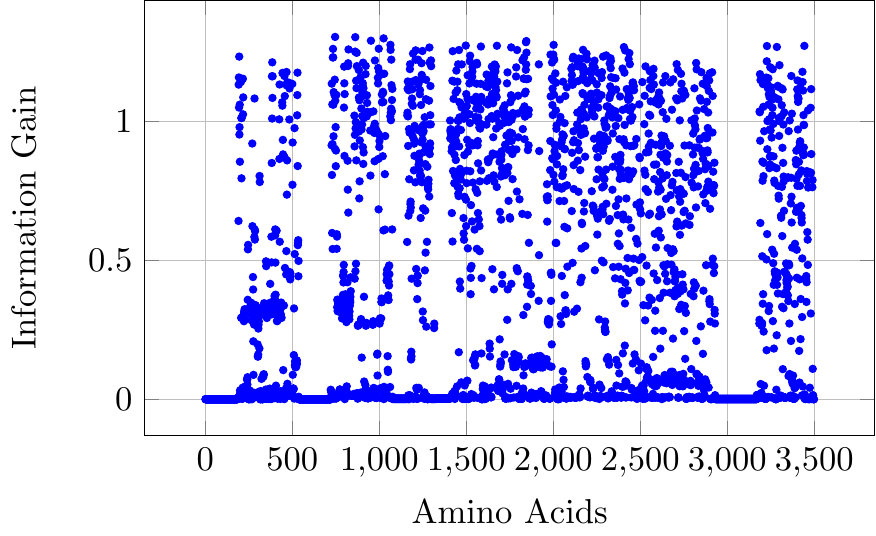}
  \caption{Information  gain  for  each  amino  acid  position  with respect  to  hosts.}
  \label{fig_data_correlation}
\end{figure}

To measure the importance of amino acids corresponding to the class label, we use the information gain (IG). 
The IG is defined as follows:
\begin{equation}
    IG(Class,position) = H(Class) - H(Class | position)
\end{equation}
\begin{equation}
    H = \sum_{ i \in Class} -p_i \log p_i
\end{equation}
where $H$ is the entropy, and $p_i$ is the probability of the class
$i$.
Figure~\ref{fig_data_correlation} shows the IG values for different amino acids corresponding to the class labels (hosts). We can see that some amino acids have higher IG values, which means that they play an important role towards the prediction of hosts. Here we can conclude that many amino acids contribute to the host specification as compared to that for SARS-CoV-2 variants specification~\cite{ali2021k}, which is expected, since the genomic variability within the family coronavidiridae should be much higher.
The IG values for all amino acids are also available online for further analysis~\footnote{\url{https://github.com/sarwanpasha/PWM2Vec/tree/main/IG\%20values}}.


Since information gain does not give us the negative (or opposite) contribution of an attribute (feature) corresponding to the class label (host names), we use other statistical measures such as Pearson correlation~\cite{benesty2009pearson} and Spearman correlation~\cite{myers2004spearman} to further evaluate the contribution of features in PWM2Vec based feature vector. The Pearson Correlation is computed using the following expression:
\begin{equation}
    r = \frac{\sum (x_i - \Bar{x}) (y_i - \Bar{y})}{\sqrt{\sum (x_i - \Bar{x})^2 (y_i - \Bar{y})^2}}
\end{equation}
where $r$ is the correlation coefficient, $x_{i}$is the values of the x-variable in a sample, $\Bar{x}$ is the mean of the values of the x-variable, $y_{i}$is the values of the y-variable in a sample, and $\Bar{y}$ is the mean of the values of the y-variable. The Spearman Correlation is computed using the following expression:
\begin{equation}
    \rho = 1 - \frac{6 \sum d_{i}^{2}}{n (n^2 - 1)}
\end{equation}
where $\rho$ is the Spearman's rank correlation coefficient, $d_{i}$ is the difference between the two ranks of each observation, and $n$is the total number of observations.

\begin{figure}[ht!]
\begin{minipage}{.5\textwidth}
\centering
\includegraphics[scale = 0.7] {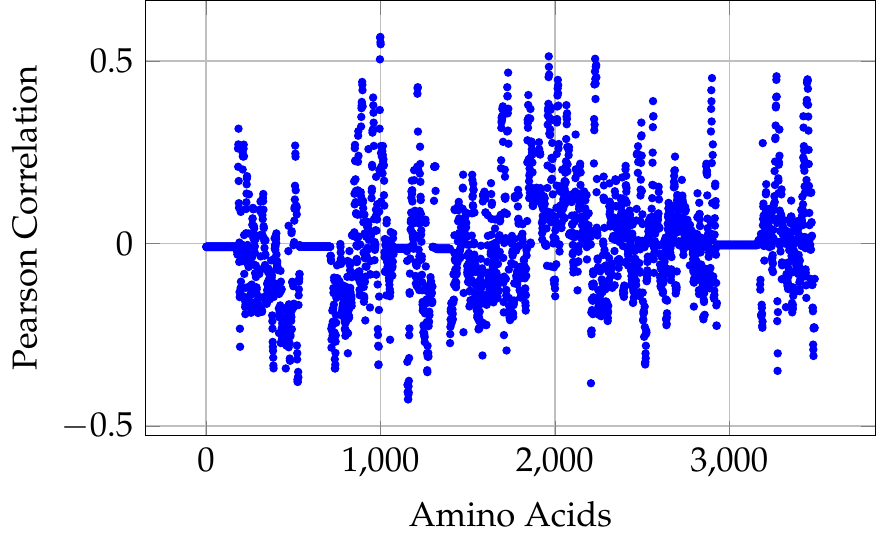}
\end{minipage}%
\begin{minipage}{.5\textwidth}
\centering
\includegraphics[scale = 0.7]{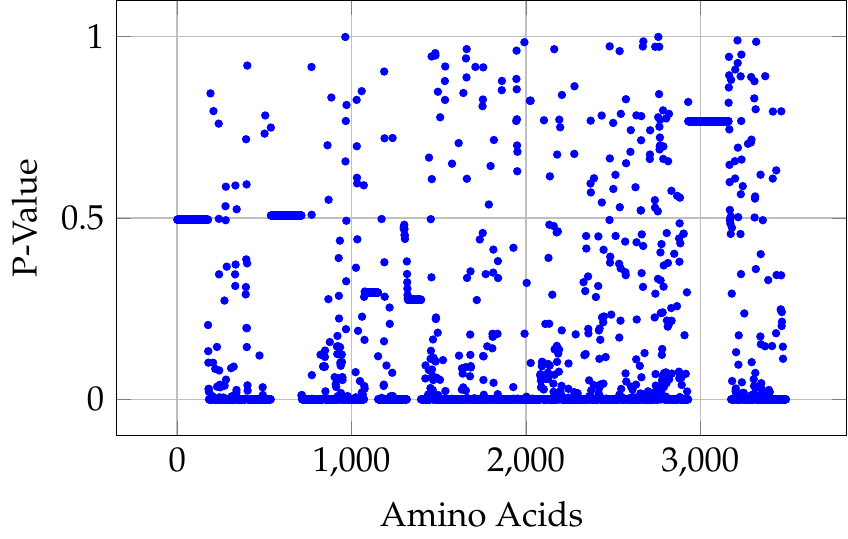}
\end{minipage}
\caption{Pearson Correlation for PWM based feature vectors.}
\label{fig_pearson_corr}
\end{figure}

\begin{figure}[ht!]
\begin{minipage}{.5\textwidth}
\centering
\includegraphics[scale = 0.7] {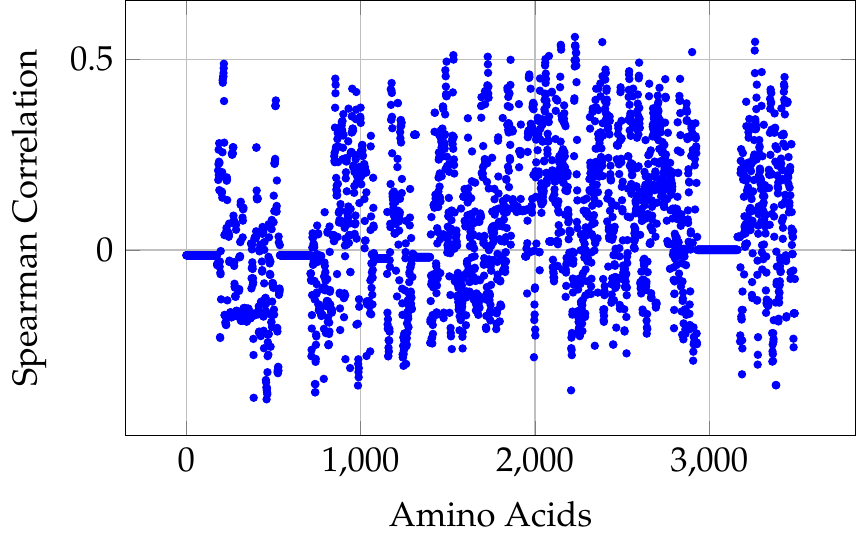}
\end{minipage}%
\begin{minipage}{.5\textwidth}
\centering
\includegraphics[scale = 0.7]{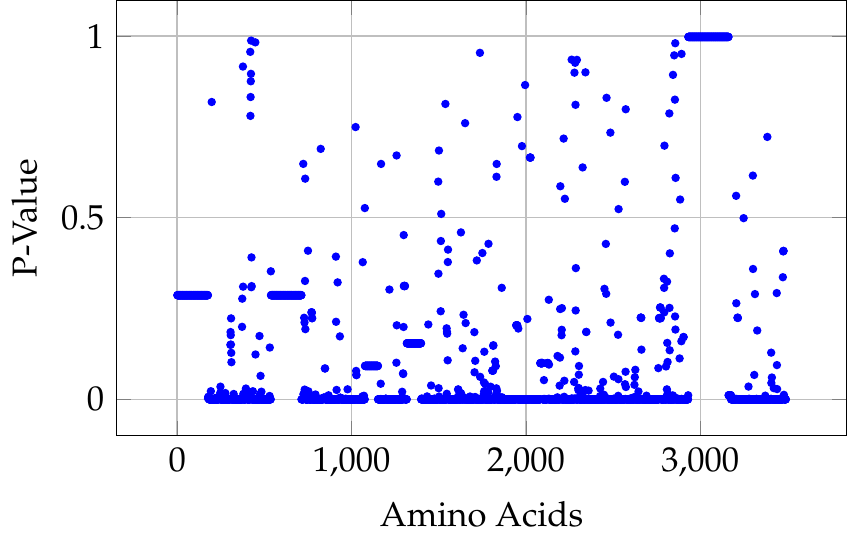}
\end{minipage}
\caption{Spearman Correlation for PWM based feature vectors.}
\label{fig_spearman_corr}
\end{figure}

The Pearson correlation and corresponding P-values for PWM2Vec are given in Figure~\ref{fig_pearson_corr} while the Spearman correlation and corresponding P-values for PWM2Vec are given in Figure~\ref{fig_spearman_corr}.  We can observe that most of the features are contributing towards the prediction of different hosts.
Table~\ref{tbl_corr_comparison} contains the comparison of correlation values computed using Pearson correlation and Spearman correlation for different embedding methods. Here, we can observe that with less dimensional and more compact approach for feature embedding (PWM2Vec), the fraction of features having correlation values greater than the threshold (i.e. $0.3$ and $-0.3$) are greater than those given for OHE and comparable with those given for k-mers (sometimes better also). This behavior shows that using PWM2Vec, we are able to preserve more information in a smaller feature vector and improve the runtime of underlying ML algorithms while giving better (sometimes comparable) predictive performance.

\begin{table}[ht!]
  \centering
  \begin{tabular}{@{\extracolsep{4pt}}lcccccccc@{}}
    \hline
    & \multicolumn{4}{c}{Pearson Correlation} & \multicolumn{4}{c}{Spearman Correlation} \\
    \cline{2-5} \cline{6-9}
    & \multicolumn{2}{c}{$>0.3$} & \multicolumn{2}{c}{$< -0.3$} & \multicolumn{2}{c}{$>0.3$} & \multicolumn{2}{c}{$< -0.3$} \\
    \cline{2-3} \cline{4-5} \cline{6-7} \cline{8-9}
    Methods & No. & frac. & No. & frac. & No. & frac. & No. & frac. \\
    \hline	\hline
    OHE & 822 & 0.009 & 1101 & 0.013 & 664 & 0.007 & 971 & 0.011 \\
    k-mers & 732 & 0.052 & 820 & 0.059 & 557 & 0.040 & 705 & 0.050 \\
    PWM2Vec & 134 & 0.038 & 51 & 0.014 & 419 & 0.120 & 33 & 0.009 \\
    \hline
  \end{tabular}
  \caption{Correlation values for different embedding approaches computed using Pearson Correlation and Spearman Correlation. We show the count (No.) and fraction (frac.) of features values greater than or less than the threshold (0.3 or -0.3). The fractions are computed by taking denominator as the size of embeddings ($83952$ for OHE, $13824$ for k-mers, and $3490$ for PWM2Vec).}
  \label{tbl_corr_comparison}
\end{table}

\section{Conclusion}\label{sec_conclusion}
We propose an approach, called PWM2Vec to generate feature vector representation for different hosts of different coronaviruses using spike sequences only, which can be used as input for different machine learning algorithms such as classification and clustering. We show that our approach is not only efficient to generate feature vectors as compared to the popular method based on $k$-mers, but has comparable prediction accuracies and much better training runtime. This behavior is also observed after applying feature selection algorithm. We also provided some statistical analysis on the data and feature vectors to show the importance of attributes towards the prediction of class labels (hosts). This statistical analysis provides a validation, from an independent point of view (in terms of fraction of features statistically correlated to the label), of the compactness of our PWM2Vec embedding, compared to the baselines.

In the future, we would focus on collecting more data to evaluate the scalability of the PWM2Vec. Using unsupervised methods for dimensionality reduction is another future extension of this work. We would also like to use deep learning models such as LSTM and GRU for the classification purposes in the future.  The application of this to larger families of viruses could also be another interesting future direction.

\reftitle{pwm}
\externalbibliography{yes}
\bibliography{pwm}

\end{document}